\documentclass[a4paper,prl,twocolumn,showpacs,superscriptaddress,floatfix,nofootinbib]{revtex4-1}
\usepackage{lipsum}
\usepackage{graphicx}
\usepackage{epsf}
\usepackage{epsfig}
\usepackage{amssymb,amsmath,amsfonts}
\usepackage[usenames]{color}
\usepackage{amssymb}
\usepackage{times}
\usepackage{bm}
\usepackage{hyperref}
\hypersetup{
  colorlinks=true,        
  linkcolor=blue,         
}

\usepackage{graphicx}  
\usepackage{dcolumn}   
\usepackage{bm}        
\usepackage{amsfonts,amsmath,amssymb,mathrsfs}
\usepackage{color}

\usepackage{hyperref}
\hypersetup{
  colorlinks=true,        
  linkcolor=blue,         
  citecolor=cyan,         
}
\usepackage{natbib,times}
\definecolor{orange}{rgb}{1,0.5,0}
\definecolor{dgreen}{rgb}{0.0, 0.5, 0.0}

\newcommand{\cf}{cf.~}
\newcommand{\ie}{i.e.,~}
\newcommand{\eg}{e.g.,~}
\newcommand{\ttt}[1]{\texttt{\small #1}}

\renewcommand{\BibitemShut}[1]{}

\newcommand{\ITP}{Institut f{\"u}r Theoretische Physik,
  Max-von-Laue-Stra{\ss}e 1, 60438 Frankfurt, Germany}
\newcommand{\FIAS}{Frankfurt Institute for Advanced Studies,
  Ruth-Moufang-Stra{\ss}e 1, 60438 Frankfurt, Germany}
\newcommand{\KENT}{Department of Physics, Kent State University, Kent, OH 44243 USA}
\newcommand{\GSI}{GSI Helmholtzzentrum f{\"u}r Schwerionenforschung GmbH, 64291 Darmstadt, Germany}

\begin{document}

\title{Signatures of quark-hadron phase transitions in general-relativistic
  neutron-star mergers}

\author{Elias~R.~Most}
\affiliation{\ITP}
\author{L.~Jens~Papenfort}
\affiliation{\ITP}
\author{Veronica Dexheimer}
\affiliation{\KENT}
\author{Matthias~Hanauske}
\affiliation{\ITP}
\affiliation{\FIAS}
\author{Stefan~Schramm}
\affiliation{\ITP}
\affiliation{\FIAS}
\author{Horst~St\"ocker}
\affiliation{\ITP}
\affiliation{\FIAS}
\affiliation{\GSI}
\author{Luciano~Rezzolla}
\affiliation{\ITP}
\affiliation{\FIAS}

\begin{abstract}
Merging binaries of neutron stars are not only strong sources of
gravitational waves, but also have the potential of revealing states of
matter at densities \emph{and} temperatures not accessible in
laboratories. A crucial and long-standing question in this context is
whether quarks are deconfined as a result of the dramatic increase in
density and temperature following the merger. We present the first fully
general-relativistic simulations of merging neutron stars including
quarks at finite temperatures that can be switched off consistently in
the equation of state. Within our approach, we can determine clearly what
signatures a quark-hadron phase transition would leave in the
gravitational-wave signal. We show that if after the merger the
conditions are met for a phase transition to take place at several times
nuclear saturation density, they would lead to a post-merger signal
considerably different from the one expected from the inspiral, that can
only probe the hadronic part of the equations of state, and to an
anticipated collapse of the merged object. We also show that the phase
transition leads to a very hot and dense quark core that, when it
collapses to a black hole, produces a ringdown signal different from the
hadronic one. Finally, in analogy with what is done in heavy-ion
collisions, we use the evolution of the temperature and density in the
merger remnant to illustrate the properties of the phase transition in a
QCD phase diagram.
\end{abstract}


\maketitle


\noindent\emph{Introduction} The 2017 detection of gravitational
waves (GWs) from a binary neutron star merger event GW170817
\cite{Abbott2017_etal} has opened a new window to study the interior of
neutron stars and, consequently, matter at nuclear densities. This
detection alone has already provided important progress on
our understanding of the maximum mass of neutron stars and on the
expected distribution in radii and tidal deformabilities
\cite{Annala2017, Margalit2017, Rezzolla2017, Radice2017, Ruiz2017,
  Shibata2017c, Most2018}. In addition, information about the merger
product itself has allowed a completely new glance on matter at densities
\emph{and} temperatures never observed or produced in a laboratory
before. To describe such environments, it is essential to employ a
microscopic model of strongly interacting matter that contains the
expected degrees of freedom and symmetries to be
present under those conditions. For this reason, we present here
neutron-star merger simulations using an equation of state (EOS) that
contains not only thermal effects but, most importantly, a description in
which the degrees of freedom change with density and temperature, going
from nucleons to hyperons and, finally, quarks.

This is substantially different from what has been done in previous
works, where either only the appearance of hyperons was investigated
\cite{Sekiguchi2011b, Radice2017a}, quarks were modeled within a
hybrid EOS (with the simple MIT bag model) at zero temperature
\cite{Oechslin:2004}, or only quarks were accounted for (again with the MIT
bag model) including temperature effects, but without a hadronic matter
component or a crust \cite{Bauswein2010} and, hence, without a phase
transition (PT). Furthermore, the last two approaches were based on a
conformally flat approximation of general relativity and evolved matter
within a smooth particle hydrodynamics (SPH) approach. In contrast, in
this \emph{Letter} we report on the first simulations employing a
self-consistent nuclear EOS including finite-temperature effects and
allowing for a first-order PT from hadrons to quarks, together with a
covariant general-relativistic description of hydrodynamics coupled to a
fully general-relativistic spacetime evolution. \\

\begin{figure*} [t!]
  \includegraphics[width=1.0\textwidth]{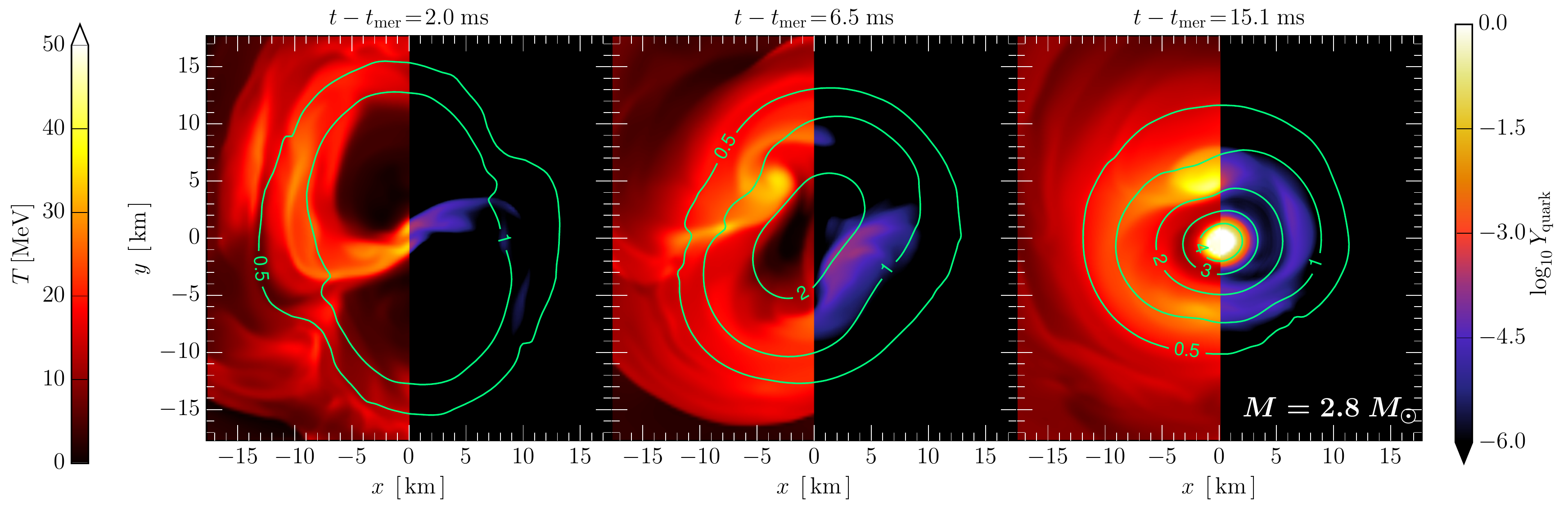}
  \caption{Snapshots on the equatorial plane at three representative
    times of the evolution of the low-mass binary. For each snapshot, the
    left part of the panel reports the temperature $T$, while the right
    part {reports} the quark fraction $Y_{\rm quark}$. The {green} lines
    show contours of constant number baryon density in units of the
    nuclear saturation density $n_{\rm sat}$. Note that a PT takes place
    only shortly before the HMNS collapses to a black hole (\cf right
    panel).}
  \label{Yquark_panel}
\end{figure*}

\noindent\emph{Methods and setup.} Matter in the inner core of neutron
stars is very dense but still strongly interacting. For this reason,
although the relevant degrees of freedom in these conditions are expected
to be quarks, first-principle theories, such as perturbative QCD, cannot
be applied directly. An alternative is to rely on effective models, which
can be calibrated to work {in the required regime of energies}. Here, we
choose the Chiral Mean Field (CMF) model, based on a nonlinear
realization of the SU(3) sigma model \cite{Papazoglou:1998}. This is an
effective quantum-relativistic model that describes hadrons and quarks
interacting via meson exchange and is constructed in a chirally-invariant
manner, since the particle masses originate from interactions with the
medium and, therefore, decrease at high densities/temperatures. The model
is in agreement with standard nuclear and astrophysical constraints
\cite{Dexheimer:2008, Negreiros:2010}, as well as lattice QCD and
perturbative QCD \cite{Dexheimer:2009, Roark:2018}. In particular, in the
limit of zero-temperature, zero-angular momentum, it predicts a maximum
mass of $2.07\,M_{\odot}$ for an hadronic star ($1.97\,M_{\odot}$ when
quarks are included) and a radius of $13.7\,{\rm km}$ for a reference
star of $1.4\,M_{\odot}$.

This approach allows for the existence of soluted quarks in the hadronic
phase and soluted hadrons in the quark phase at finite
temperature. However, quarks/hadrons will always give the dominant
contribution in the quark/hadron phase, and the two phases can be
distinguished through their order parameters. This inter-penetration of
quarks and hadrons (that increases with temperature) provides a
physically effective description and is indeed required to achieve the
crossover transition known to take place at small chemical potential
values \cite{Aoki:2006}. While this approach is suitable to describe
matter in the neutron-star core, another description is needed for the
crust and the very low density regions produced in binary mergers. For
these, we have matched the CMF EOS to the nuclear statistical
equilibrium description presented in ~\cite{daSilvaSchneider2017}.

To describe the evolution of the merging system, we solve the coupled
Einstein-hydrodynamics system \cite{Rezzolla_book:2013} using the newly
developed \ttt{Frankfurt/IllinoisGRMHD} code (\ttt{FIL}), which is a
high-order extension of the publically available \ttt{IllinoisGRMHD} code
\citep{Etienne2015}, part of the \ttt{Einstein Toolkit}
\citep{loeffler_2011_et}. In particular, \ttt{FIL}, which belongs to the
family of Frankfurt Relativistic-Astrophysics Codes (\ttt{FRAC}),
implements a fourth-order accurate conservative finite-difference scheme
\citep{DelZanna2007} using a WENO-Z reconstruction \citep{Borges2008},
coupled to an HLLE Riemann solver \cite{Harten83}. The code handles
temperature dependent EOSs utilizing a novel infrastructure, and the
conversion from conservative to primitive variables follows
~\cite{Galeazzi2013} for purely hydrodynamical simulations. To account
for weak interactions, a neutrino-leakage scheme is implemented following
~\cite{Ruffert96b, Rosswog:2003b, OConnor10}. The \ttt{FIL} code can also
handle neutrino heating via an M0 scheme \cite{Radice2016} and has
recently participated in a multi-group code comparison demonstrating its
ability to provide an accurate and fully convergent description of the
dynamics of merging compact stars. We will comment further on the its
capabilities in an upcoming publication.

The spacetime is evolved using the Z4c formulation of the Einstein
equations \citep{Bernuzzi:2009ex}, which is a conformal variant of the Z4
family \cite{Bona:2003fj} (see also \cite{Alic:2011a}), following the
setup in \citep{Hilditch2012}, while the gauges are the same as in
\cite{Hanauske2016,Bovard2017}. The initial data is modeled under the
assumption of irrotational quasi-circular equilibrium
\cite{Gourgoulhon-etal-2000:2ns-initial-data} and is computed by the
LORENE library. The binaries are initially at a distance of $45 \,{\rm
  km}$ and perform around five orbits before the merger. The numerical
grid uses the fixed-mesh refinement driver \ttt{Carpet}
\cite{Schnetter-etal-03b}, with a total of seven refinement levels having
a highest resolution of $\simeq 250\, \rm m$ covering the two stars and a
total extent of $\simeq 1500\, \rm km$. 

\noindent\emph{Results.} While we have evolved a larger spectrum in
masses for binaries with either equal or unequal masses, we next
concentrate on two cases that best illustrate the onset of a first-order
PT. These are equal-mass binaries with total masses $M=2.8$ and $2.9\,
M_\odot$, hereafter referred to as the low- and high-mass binaries,
respectively. Lower-mass binaries lead to post-merger objects with zero
or minute quark fraction, while higher-mass binaries collapse to a black
hole before a PT can fully develop. As anticipated above, a distinctive
feature of our approach is the ability to cleanly and robustly determine
the role of quarks in the merger remnant by using the same EOS with and
without quarks. Because of this, for each of the two masses we perform
two identical simulations either employing the standard CMF EOS where
quarks and a strong first-order PT are included (\ie ${\rm CMF}_{\rm
  Q}$), or a purely hadronic version in which the quarks are not included
(\ie ${\rm CMF}_{\rm H}$). In the case of the high-mass, ${\rm CMF}_{\rm
  Q}$ binary, we have also performed a simulation with a very-high
resolution of $\simeq 125\, \rm m$. Leading only to a $1.5\%$ difference in
the collapse time, this confirms that the reference resolution reported
here is sufficient to capture qualitatively the dynamics of the PT.

We begin by describing the overall evolution during and after the merger
of the low-mass binary with total mas $M=2.8\,M_{\odot}$. In particular,
Fig. \ref{Yquark_panel} reports three representative snapshots on the
equatorial plane. Right after the merger time $t_{\rm mer}$, and slightly
before the time shown in the left panel of Fig. \ref{Yquark_panel}, the
regions with high temperatures are near the central regions of the
hypermassive neutron star (HMNS). Some time later, and in analogy with
what was shown in previous studies \cite{Kastaun2016, Hanauske2016}, the
temperature distribution shows two ``hot spots'' in spatially opposite
regions (middle panel) that also correspond to local minima of the number
density (see \cite{Hanauske2016} for a detailed discussion in terms of
the Bernoulli constant). Interestingly, already a few milliseconds after
the merger, a small but nonzero amount of quarks constituting $\lesssim
0.02\%$ of the total baryon mass begins to appear in regions of high
temperature and \emph{before} a first-order PT occurs. Because even small
fractions of quarks can alter the pressure, the quadrupole moment of the
HMNS will be different when compared to the pure hadronic case. As time
progresses, the hot spots merge {into} a ring (right panel), at which
time also the density has reached the critical value for the onset of the
PT, leading to the production of a large amount of quarks in the core of
the HMNS. When this happens, the quark fraction $Y_{\rm quark}$ can be as large
as $0.9$ locally and quarks represent $\sim 15\!-\!20 \%$ of the total baryon
mass.

\begin{figure} [t!]
  \includegraphics[width=1.0\columnwidth]{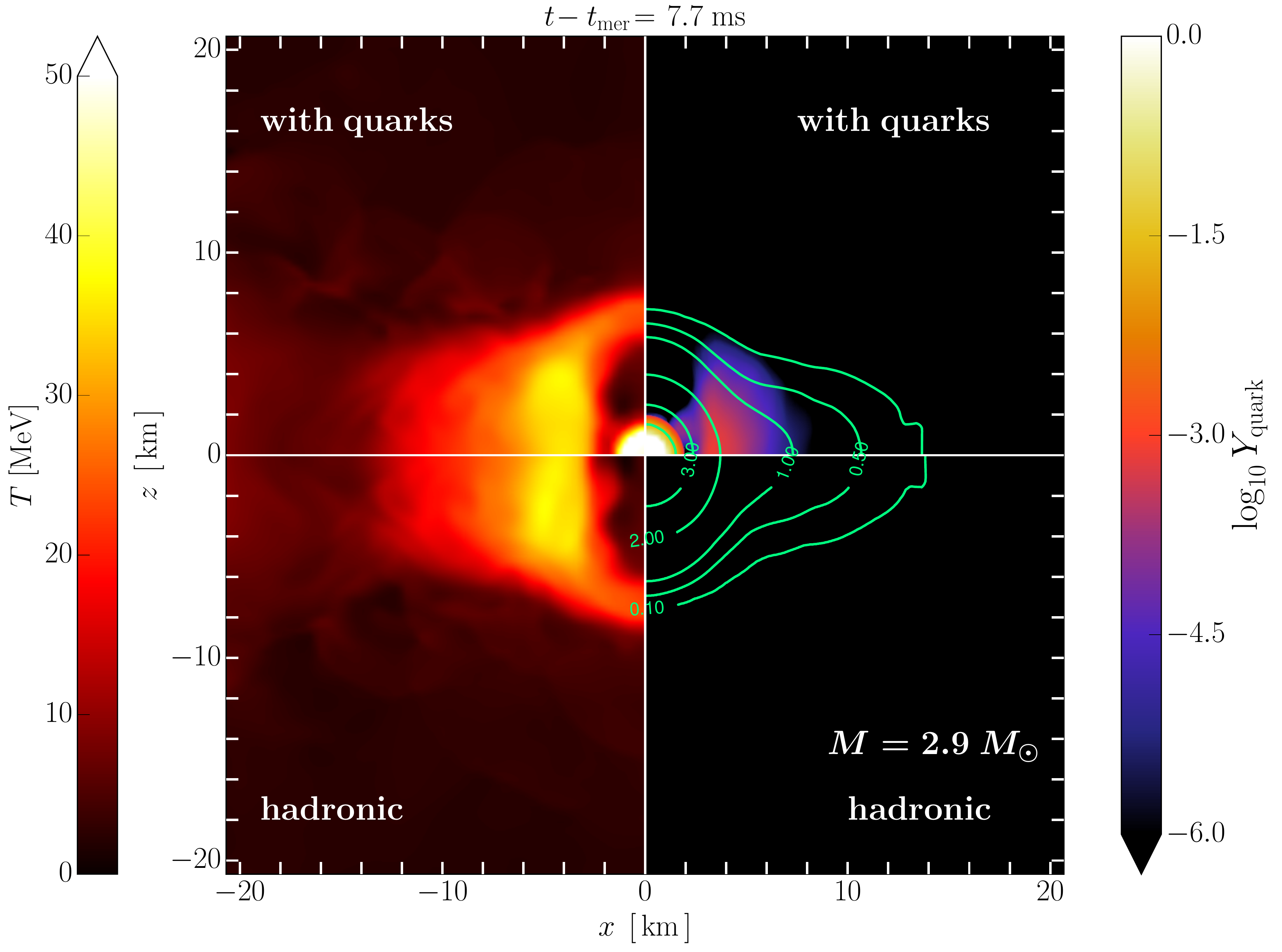}
  \caption{Same as Fig. \ref{Yquark_panel} but on the meridional plane;
    the top/bottom panels show simulations with the ${\rm CMF}_{\rm
      Q}$/${\rm CMF}_{\rm H}$ EOSs, respectively. The snapshot refers to
    the high-mass binary at a time shortly before the collapse to a black hole.}
  \label{fig:CMF_rhoT}
\end{figure}

The onset of the PT has a dramatic effect on the equilibrium of the
HMNS. The very rapid softening of the EOS, in fact, leads to a rapid
compression of the central region of the HMNS; the resulting release of
gravitational binding energy produces a sharp increase in the baryon
number density and a massive heat-up of the core that, in the absence
of the PT, would be cold.

To appreciate this better, Fig. \ref{fig:CMF_rhoT} shows the same
quantities as Fig. \ref{Yquark_panel}, but in the meridional plane for
the high-mass binary and after the PT has taken place. Different panels
compare simulations performed with the ${\rm CMF}_{\rm Q}$ EOS, where
quarks are present (top panels) with simulations employing the ${\rm
  CMF}_{\rm H}$ EOS, in which quarks are suppressed (bottom panels). It
is remarkable to note the large quark fraction in the center and also in
regions of high temperature (top-right panel), which is, of course,
absent for the ${\rm CMF}_{\rm H}$ EOS (bottom-right panel). Similarly,
while the temperature distributions are very similar in the outer parts
of the HMNS, where the densities are comparatively low, they are very
different in the inner regions.

Finally, it should be emphasized that the ${\rm CMF}_{\rm Q}$ EOS does
not lead to the formation of a gravitationally stable quark phase and,
therefore, the very massive quark core collapses essentially in free
fall, \ie in $\lesssim 1\ \rm ms$, to a rotating black hole. As discussed
in ~\cite{Hempel:2013}, a relaxation of the charge-neutrality
constraint in the EOS from being local to being global (a so-called Gibbs
construction) would create a stable mixture of phases that, in the case
of massive and isolated stars, would extend to several kilometers within
the star. We here do not relax such constraint as we are interested in
studying the effect of a steep first-order PT and thus in finding the
most extreme signals that could be produced in such events.

\begin{figure} [t!]
  \includegraphics[width=1.0\columnwidth]{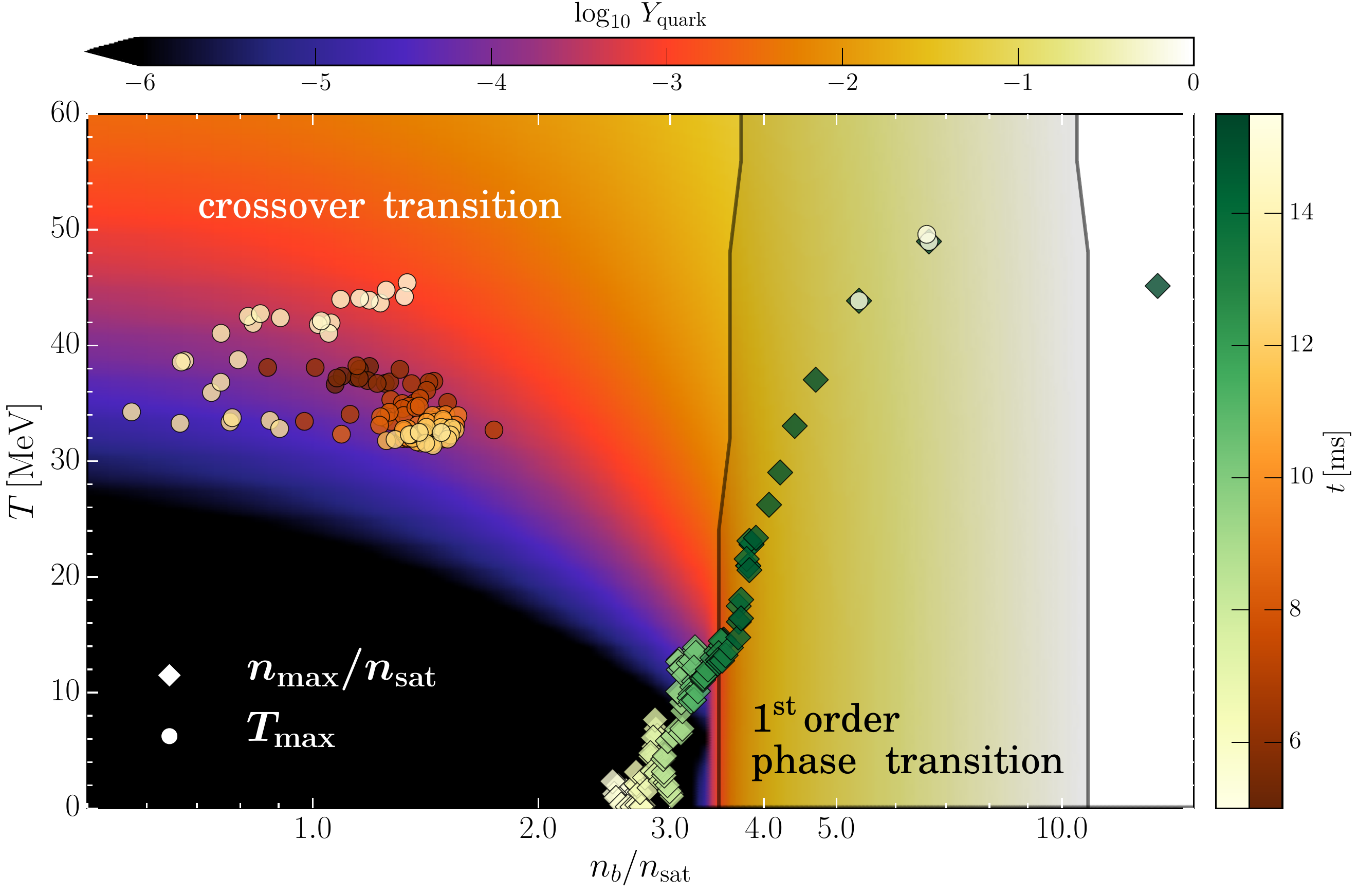}
  \caption{Evolution of the maximum normalized baryon number density
    (diamonds) and temperature (circles) after the merger for the
    low-mass binary with the ${\rm CMF}_{\rm Q}$ EOS. Different times of
    the evolution are represented with a color code, together with the
    quark fraction $Y_{\rm quark}$.
    The grey shaded area shows the first-order PT region.}
  \label{fig:CMF_rhoT_evol}
\end{figure}

\begin{figure*}
  \includegraphics[width=1.0\textwidth]{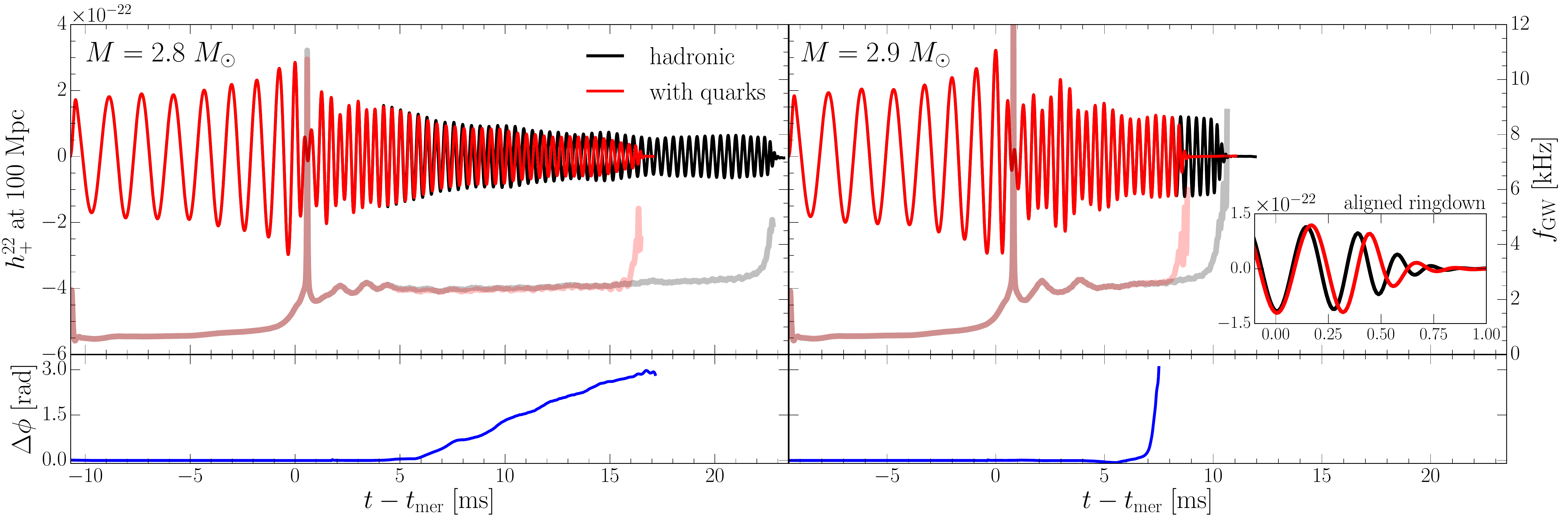}
  \caption{Properties of the GW emission for the low- (left panels) and
    high-mass binaries (right panels). The top panels report the strain
    $h^{22}_+$
    for the two EOSs, together with the instantaneous
    GW frequency $f_{\rm GW}$ (semitransparent lines); the bottom panels show the
    phase difference $\Delta \Phi$ between the two signals. The inset in the top-right
    panel highlights the differences in the ringdown.}
  \label{fig:strain}
\end{figure*}

Since the deconfinement of quarks depends on both the density and the
temperature of matter, it is interesting to consider which regions of the
EOS are actually probed by the merger remnant. Hence, in close analogy
with what is done in heavy-ion collisions \cite{Foka2017},
Fig.~\ref{fig:CMF_rhoT_evol} shows the evolution of the maximum baryon
number density $n_{\rm max}$ (normalized to the nuclear saturation
density $n_{\rm sat}$) and of the maximum temperature $T_{\rm max}$ for
the low-mass binary evolved with the ${\rm CMF}_{\rm Q}$ EOS. The
time-series spans a time between five and $15 \, \rm ms$ after merger. In
essence, diamonds refer to the part of the matter in the core of the
HMNS, while circles illustrate the conditions in the hot and low-density
regions affecting the quadrupole moment. Note that matter in the HMNS
core exhibits higher densities of $> 2 \, n_{\rm sat}$, but also
temperatures below $10 \, \rm MeV$; by contrast, heavy-ion collisions of,
\eg ${\rm Au} + {\rm Au}$ at energies of $\sim 0.5-1\, {\rm GeV}$, probe
number densities above $n_{\rm sat}$ and temperatures of $50 \lesssim
T/{\rm MeV} \lesssim 100$ (not shown in Fig.~\ref{fig:CMF_rhoT_evol}).

Loss of angular momentum through GWs leads to a continuous rise of the
central density (hence of $n_{\rm max}/n_{\rm sat}$), which ultimately
reaches the boundary of the first-order PT (grey-shaded area) in
Fig.~\ref{fig:CMF_rhoT_evol} at $\sim 13 \, \rm ms$ after merger. While
contracting, the core of the HMNS crosses this region very rapidly and
establishes an almost-pure quark phase heated up to temperatures $>40 \,
\rm MeV$. If metastable, this core might influence the surrounding
material, although the densities inside the HMNS are so high that
neutrinos are essentially trapped.  As can be seen from the last marker
of the density evolution in Fig. \ref{fig:CMF_rhoT_evol}, the HMNS core
undergoes a complete PT to quarks and the whole HMNS collapses
immediately after the PT. Note that the region of highest temperature is
initially at densities smaller than $\sim n_{\rm sat}$, but the temperature is
sufficiently high for quarks to appear in small amounts. After the HMNS core
crosses the PT boundary, the maximum temperature rises steeply and thus
the fluid elements with maximum density and temperature coincide.

We complete our discussion of the PT by considering its signatures on the
GW emission by means of the strain, frequency and phase difference, which
are reported in Fig. \ref{fig:strain} for the low- and high-mass
binary. Note that because the densities and temperatures during the
inspiral are too small to cause the formation of quarks, the GW signal is
identical for the two EOSs and for both masses. This is radically
different from what happens when comparing merger simulations using EOSs
with and without hyperons, as these show differences in the GW signal
already during the inspiral \cite{Sekiguchi2011b, Radice2017a}, due to the
softening caused by the presence of hyperons. For such EOSs, a dephasing is
thus \emph{always} present, both during the inspiral and after the merger,
since there are always portions of the stars with intrinsically different
EOSs. In our case, instead, it is only \emph{after} the merger that
differences arise due to the presence of quarks.

For the low-mass binary, and after $\sim 5\,\rm ms$ from the merger, the
GWs from the remnants start to show a linear dephasing that reaches about
three radians by the time the binary with the ${\rm CMF}_{\rm Q}$ EOS
collapses to a black hole (bottom-left panel). The start of the phase
difference, which is essentially zero even after the merger, coincides
with the formation of the two hot spots and, thus, with the appearance of
quarks. In fact, although $Y_{\rm quark}$ is very small at those times,
it is sufficient to produce changes in the pressure of $\sim 5\%$, that
are responsible for the changes in the GW emission, both in amplitude and
in frequency (top-left panel), thus producing a mismatch between two
post-merger spectra \cite{Bauswein2011, Stergioulas2011b, Takami2014,
  Takami2015, Bernuzzi2015a, Rezzolla2016}. These changes in pressure
also lead to a small damping of the GW amplitude prior to collapse, which
is triggered by the first-order PT for the ${\rm CMF}_{\rm Q}$
EOS. Hence, the lifetime of the HMNS is shorter than in the purely
hadronic case.

In many respects, the left panels of Fig. \ref{fig:strain} are a
representative example of the signatures of a PT in a binary merger. In
an idealized scenario where the GW signal from the inspiral is measured
with great precision and can be associated with confidence to a purely
hadronic EOS (the inspiral can only probe comparatively low-density
regions of the EOS), the unexpected dephasing of the template-matched
post-merger signal \cite{Bose2017, Chatziioannou2017}, together with the
anticipated collapse of the HMNS, would provide evidence that a PT at
several times $n_{\rm sat}$, possibly of the type described here, has
taken place in its core. Of course, a single detection could still be
accomodated via a tweaking of the EOS in the high-density part of a
hadronic EOS. However, the ``tweaking'' would be increasingly hard with
multiple detections as it cannot describe the complex nonlinear
occurrence of the PT.  

The right panels of Fig. \ref{fig:strain} report the properties of the GW
signal for the high-mass binaries, both of which collapse very rapidly
for EOSs with and without quarks. The differences in this case are harder
to detect since the dephasing starts only after $\sim 5\,{\rm ms}$, but
is very quickly suppressed by the collapsing signal. The latter, however,
is different, as shown in the small inset in the top-right panel of
Fig. \ref{fig:strain}, where the two ringdown signals are suitably
aligned. These differences are caused by distinct free-fall times of the
cores of the HMNSs, which are shorter in the case of the ultra-softened
EOS with quarks. Although these differences are not large (the relative
difference in the ringdown-frequency is $\lesssim 25 \%$, yielding an
overlap of only $\mathcal{O}=0.92$ \cite{Lindblom08,Giacomazzo:2009mp})
they are large enough to be distinguishable if detected by
third-generation GW detectors \cite{Punturo:2010, Evans2016}.  As a final
remark, we point out that all of the dynamics reported above is found
also when simulating unequal-mass binaries with mass ratio $q=0.8$; the
main difference in this case is that the PT occurs off-center because the
high-density region is also off-centered.

\noindent\emph{Conclusions.} We have presented the first fully
general-relativistic simulations of merging binary neutron stars
including quarks at finite temperatures. Because in our approach the
presence of quarks can be turned off consistently, it was possible to
study their imprint on the merger in a clean and robust manner. Moreover,
since our description allows for the appearance of small amounts of
quarks below the PT region related to the crossover transition at low
densities, we were able to observe significant nonzero quark fractions in
regions of high temperature but densities below saturation. The changes
in pressure produced by these soluted quarks were shown to lead to a
systematic dephasing \textit{only} of the post-merger GW emission, which,
if accumulated over several milliseconds, can produce a decisive
signature in the post-merger GW signal and spectrum. This behaviour is
markedly different from that of other hyperonic EOSs, which show softenings
already during the inspiral. Furthermore, the inclusion of a first-order
PT in a thermodynamically consistent way has allowed us to associate the
PT with the formation of a very hot and ultra-dense quark core in the
HMNS, that was gravitationally unstable and collapsed to a black
hole. Finally, despite the short lifetime of the quark phase, we have
shown that its collapse, which proceeds essentially in free-fall, leads
to different black-hole ringdown frequencies, another useful signature of
the occurrence of the PT.

The work presented here can be extended in at least three ways. Firstly,
by considering an EOS that would allow for the existence of a metastable
quark core after a PT. In this case, the post-merger GW spectrum would be
the combination of the spectrum of the purely hadronic remnant together
with that produced after the PT. Secondly, a PT to a metastable quark
core in the HMNS could lead (either immediately or after a diffusion
timescale) to a burst signal in neutrinos in analogy with what suggested
for supernovae \cite{Sagert2011, Fischer2017}, providing yet another
evidence of the PT. Finally, the occurrence of the PT will also impact
the spectral properties of the post-merger signal to the point that the
$f_2$ oscillation frequency may exhibit a jump to higher frequency as a
result of the PT to a different metastable HMNS \cite{Hanauske2017,
  Hanauske2018}. These scenarios will be explored in future works.

\medskip
It is a pleasure to thank M. Alford, T. Galatyuk, J. Schaffner-Bielich,
J. Steinheimer, and J. Stroth for useful discussions. Support comes also
in part from ``PHAROS'', COST Action CA16214; LOEWE-Program in HIC for
FAIR; European Union's Horizon 2020 Research and Innovation Programme
(Grant 671698) (call FETHPC-1-2014, project ExaHyPE); the ERC Synergy
Grant ``BlackHoleCam: Imaging the Event Horizon of Black Holes'' (Grant
No. 610058), and by the National Science Foundation under grant
PHY-1748621. The simulations were performed on the SuperMUC cluster at
the LRZ in Garching, on the LOEWE cluster in CSC in Frankfurt, and on the
HazelHen cluster at the HLRS in Stuttgart.


\bibliography{aeireferences} \bibliographystyle{apsrev4-1}

\end{document}